\documentclass[aps,prl,twocolumn,superscriptaddress]{revtex4-1}
\usepackage{graphicx,color,graphics}
\usepackage{dcolumn} 
\usepackage{amsmath}
\usepackage{amssymb}
\usepackage{amsfonts}
\usepackage{url}
\usepackage{bm}
\usepackage{lipsum}
\usepackage{xcolor}
\makeatletter

\newcommand{\ket}[1]{{{|}{#1}\rangle}}

\newcommand\blankfootnote[1]{%
  \let\thefootnote\relax\footnotetext{#1}%
  \let\thefootnote\svthefootnote%
}

\newif\ifcmnt
\cmnttrue

\ifcmnt
    \providecommand{\aucmnt}[1]{#1}
\else
    \providecommand{\aucmnt}[1]{}
\fi

\begin{document}

\title{Suppressing inhomogeneous broadening in a lutetium multi-ion optical clock.}
\author{T. R. Tan}
\affiliation{Centre for Quantum Technologies, National University of Singapore, 3 Science Drive 2, 117543 Singapore}
\affiliation{Department of Physics, National University of Singapore, 2 Science Drive 3, 117551 Singapore}
\author{R. Kaewuam}
\affiliation{Centre for Quantum Technologies, National University of Singapore, 3 Science Drive 2, 117543 Singapore}
\author{K. J. Arnold}
\affiliation{Centre for Quantum Technologies, National University of Singapore, 3 Science Drive 2, 117543 Singapore}
\author{S. R. Chanu}
\affiliation{Centre for Quantum Technologies, National University of Singapore, 3 Science Drive 2, 117543 Singapore}
\author{Zhiqiang Zhang}
\affiliation{Centre for Quantum Technologies, National University of Singapore, 3 Science Drive 2, 117543 Singapore}
\author{M. S. Safronova}
\affiliation{Department of Physics and Astronomy, University of Delaware, Newark, Delaware 19716, USA}
\affiliation{Joint Quantum Institute, National Institute of Standards and Technology and the University of Maryland,
College Park, Maryland, 20742}
\author{M. D. Barrett}
\email{phybmd@nus.edu.sg}
\affiliation{Centre for Quantum Technologies, National University of Singapore, 3 Science Drive 2, 117543 Singapore}
\affiliation{Department of Physics, National University of Singapore, 2 Science Drive 3, 117551 Singapore}
\date{\today}

\begin{abstract}
We demonstrate precision measurement and control of inhomogeneous broadening in a multi-ion clock consisting of three $^{176}$Lu$^+$ ions. Microwave spectroscopy between hyperfine states in the $^3D_1$ level is used to characterise differential systematic shifts between ions, most notably those associated with the electric quadrupole moment.  By appropriate alignment of the magnetic field, we demonstrate suppression of these effects to the $\sim 10^{-17}$ level relative to the $^1S_0\leftrightarrow{}^3D_1$ optical transition frequency. Correlation spectroscopy on the optical transition demonstrates the feasibility of a 10\,s Ramsey interrogation in the three ion configuration with a corresponding projection noise limited stability of $\sigma(\tau)=8.2\times 10^{-17}/\sqrt{\tau}$. 
\end{abstract}
\maketitle

With fractional uncertainties near to $\sim 10^{-18}$, state-of-the-art optical atomic clocks are among the most accurate scientific artefacts \cite{Ludlow2015}. The two most successful realizations are ensembles of neutral atoms stored in optical lattices \cite{Nicholson2015,McGrew2018} and ions confined in radio-frequency (RF) traps \cite{Chou2010,Huntemann2016}. The latter offers strong confinement such that atoms can be reused for subsequent clock interrogation and measurement. The stability of the current generation of trapped-ion optical clocks is limited by single-ion operation. This has limited the instability of ion-based clocks to $\sim 10^{-15}/\sqrt{\tau}$, for which averaging times $\tau$ of several days or even weeks are required to reach $10^{-18}$ resolution.  Modest improvements to stability can be expected as laser technology develops to allow longer interrogation times but ideally this would go hand-in-hand with an increase in the number of ions.

Within the standard quantum limit (SQL), clock stability improves with the $\sqrt{N}$ where $N$ is the number of atoms \cite{Itano1993}. With an ensemble of ions, frequency resolution could be further enhanced using entangled states \cite{Wineland1994,Andre2004,Bollinger1996,Leibfried2004,Shaniv2018} or cascaded interrogation schemes \cite{Borregaard2013,Rosenband2013arXiv}. From a technological standpoint, extension of clock operation to a small ensemble of ions is an immediate application for devices developed for small-scale quantum information processing (QIP).  However, characterizing and maintaining exquisite control over various systematic effects in an ion ensemble is a significant challenge. 

Multi-ion operation is complicated by electric quadrupole (EQ) shifts arising from the Coulomb fields of neighbouring ions, excess-micromotion (EMM) shifts induced by the radio-frequency (rf) trapping field, and inhomogeneous magnetic fields. Efforts and proposals towards high-accuracy multi-ion optical clocks include (i) precision engineering of the ion trap to suppress EMM shifts \cite{Keller2019}, (ii) employing clock transitions with a negative differential scalar polarisability, $\Delta\alpha_0$, to eliminate EMM shifts in a large ion crystal \cite{Berkeland1998,Arnold2015}, and (iii) using dynamic decoupling or rf-dressed states to suppress EQ shifts \cite{Shaniv2018arXiv,aharon2018robust}.  All three approaches are readily adaptable to $^{176}$Lu$^+$ although the first approach requires EQ shifts to be mitigated in a linear ion crystal.  

In this work, high resolution microwave spectroscopy is used to characterize the effects of neighbouring ions in a three-ion crystal.  This facilitates the alignment of the magnetic field to suppress ion-induced EQ shifts to the few mHz level, or $\sim 10^{-17}$ relative to the $^1S_0\leftrightarrow{}^3D_1$ optical transition frequency.  The resulting shift of the clock transition would be further suppressed by hyperfine averaging \cite{Barrett2015}.  Correlation spectroscopy, which allows frequency comparison between ions beyond the coherence time of the probe laser, is used to demonstrate multi-ion atomic coherence for interrogation times up to 10\,s.

\begin{figure}
\includegraphics[width=0.65\linewidth]{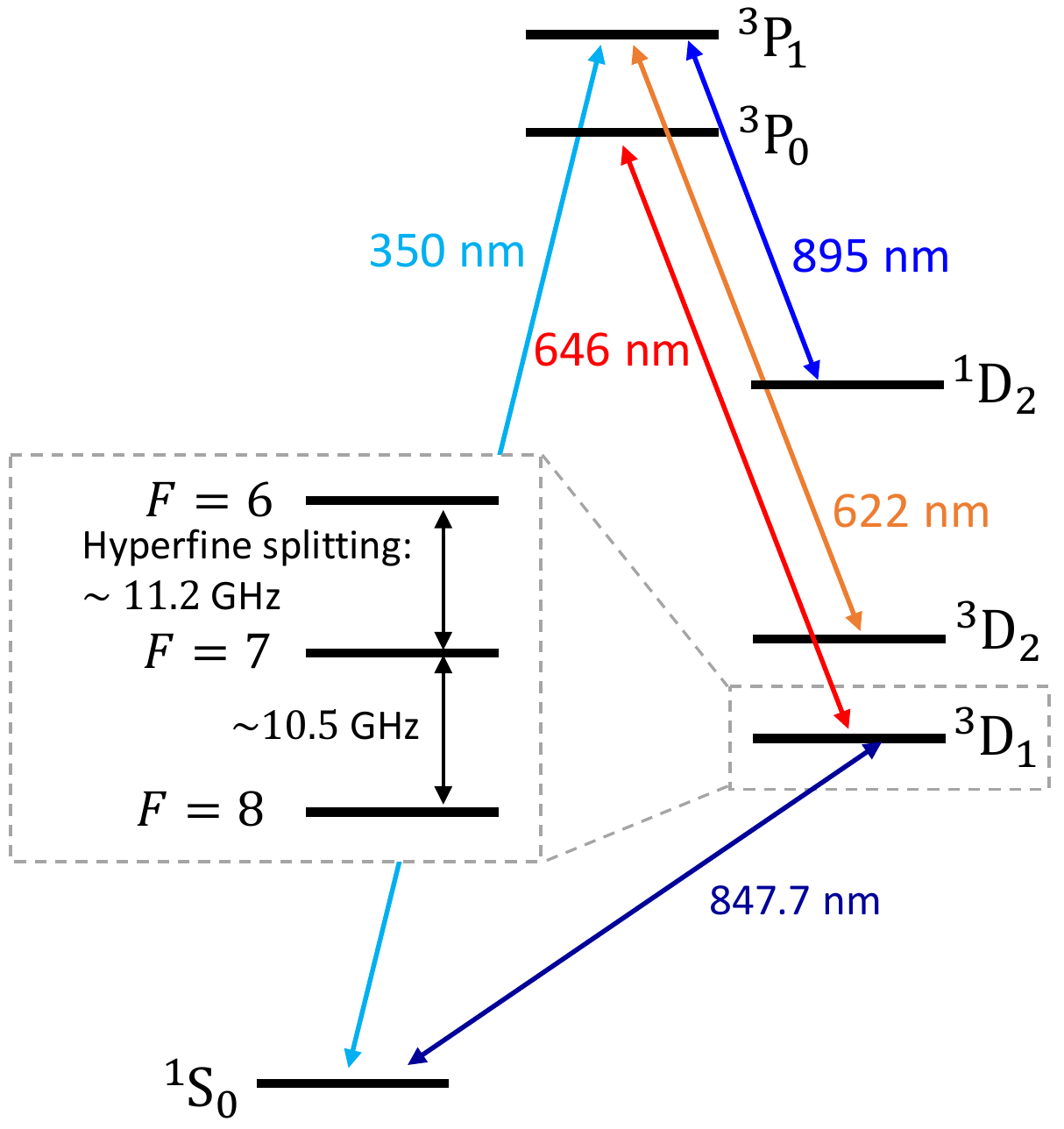}
\caption{Relevant energy level diagram of a $^{176}$Lu$^+$ ion showing: the $^1S_0 \leftrightarrow {}^3D_1$ clock transition at 848\,nm, the $^3D_1 \leftrightarrow {}^3P_0$ transition at 646\,nm used for Doppler cooling and detection, and optical pumping transitions at 350 nm, 622 nm, and 895 nm to initialize population to the $^3D_1$ level.}
\label{LuEnergyLevel}
\end{figure}

\begin{figure}
\includegraphics[width=0.68\linewidth]{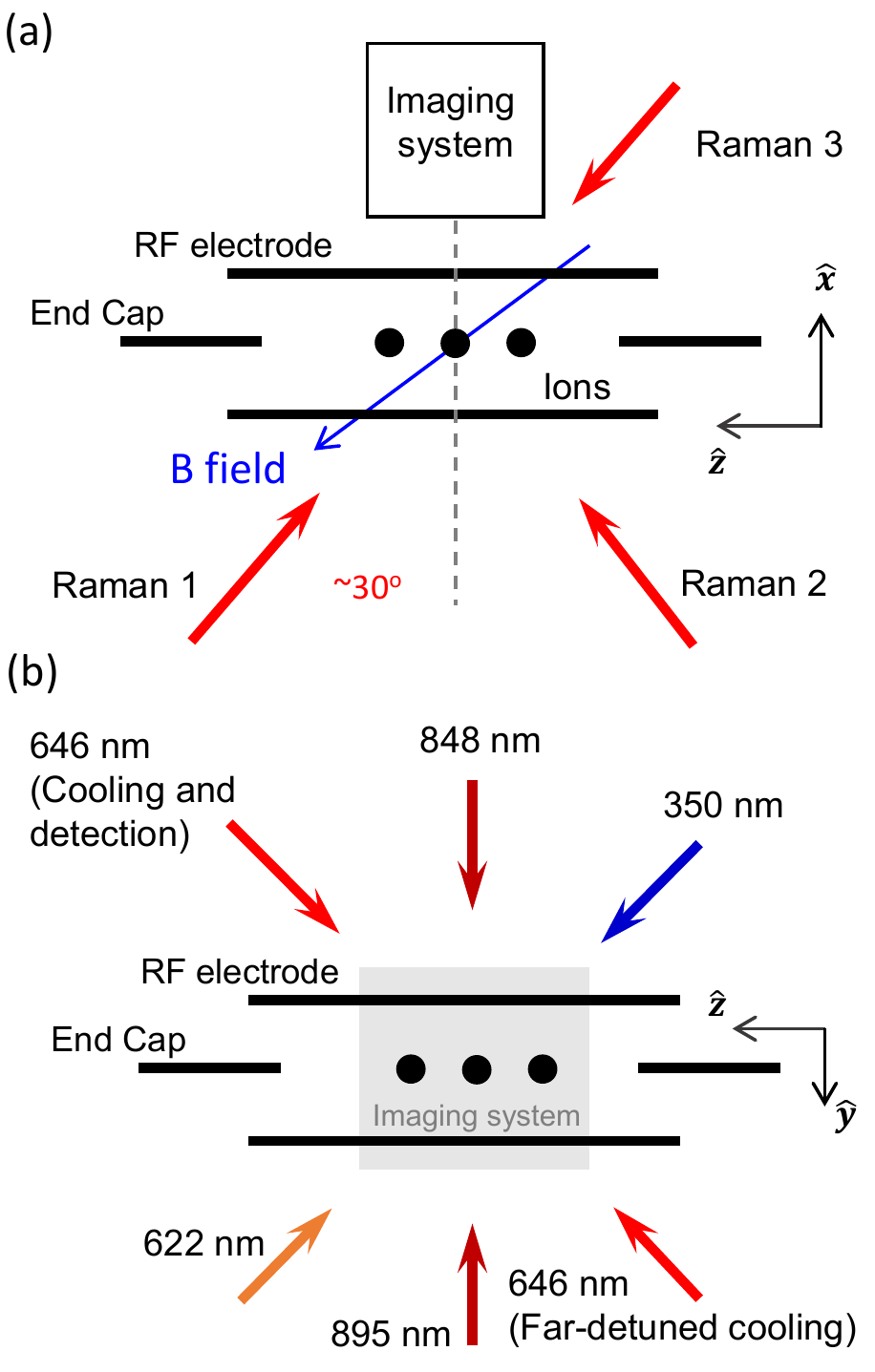}
\caption{Schematic of the experimental set up showing laser beam configurations relative to the trap. (a)\&(b) show views along the $y$ (vertical) and $x$ (horizontal) axes, respectively.  The B-field defines the quantization axis and is rotated in the $x-z$ plane to minimize tensor shifts. Raman beams in (a) and a near-resonant 646-nm beam with $\bm{k}$-vector of $\frac{1}{\sqrt{2}}(\hat{y}-\hat{z})$ in (b) are used to assess EMM. } 
\label{TrapAndBeamSetup}
\end{figure}

Lutetium offers three possible clock transitions, with the focus here being on the $^1S_0\leftrightarrow{}^3D_1$ transition.  An energy level diagram relevant to this work is shown in Fig. \ref{LuEnergyLevel}. Optical pumping with 350, 622, and 895\,nm laser beams initialize population in the $^3D_1$ level.  Doppler cooling and fluorescence measurements are accomplished with 646\,nm laser beams addressing the $^3D_1\leftrightarrow{}^3P_0$ transition, which has a linewidth of $2\pi \times 2.45$ MHz \cite{Kaewuam2017}.  An additional $\pi$-polarized 646-nm beam addressing $F=7$ to $F'=7$ facilitates state preparation into $\ket{{}^3D_1,7,0}$.  All beams are switched with acousto-optic modulators (AOM), and mechanical shutters are placed in 646- and 350-nm beam lines to extinguish leakage light arising from the finite extinction ratio of the AOMs. Laser configurations relative to the trap are as shown in Fig.~\ref{TrapAndBeamSetup}.

Experiments are carried out in a four-rod linear Paul trap with axial end caps as described elsewhere \cite{Kaewuam2017,Kaewuam2019}. An ac field of frequency $\Omega_\mathrm{rf}=2\pi\times16.74\,\mathrm{MHz}$ is applied to diagonally opposing electrodes via a quarter wave resonator and dc voltages of $\sim 8\,\mathrm{V}$ are applied to the end-caps.  This configuration provides measured secular frequencies of $\{\omega_{r_1},\omega_{r_2},\omega_z\}=2\pi \times \{720, 680, 133\}\,\mathrm{kHz}$.  A magnetic field of $\sim 0.1\,\mathrm{mT}$ is used to lift the degeneracy of Zeeman levels. Microwave transitions between $^3D_1$ hyperfine levels are driven by an antenna located outside the vacuum chamber.  

For detection, 646-nm fluorescence is imaged onto an EMCCD camera with a pixel size of $16\,\mathrm{\mu m}$.  The imaging system has an estimated magnification factor of $\sim 7$ and a numerical aperture of $0.4$.  An algorithm based on a maximum likelihood estimator is used to minimize crosstalk between adjacent ions and distinguish the eight possible outcomes. By applying the algorithm on reference images, the overall detection error is estimated to be 0.5\%.

For Lu$^+$, a technique of hyperfine averaging can be used to eliminate shifts associated with the electronic angular momentum, $J$, which appear differentially between hyperfine states, $\ket{F,m_F}$, of a $J>1/2$ fine-structure level \cite{Barrett2015}.  However, in a multi-ion crystal, these shifts are a source of inhomogeneous broadening, which would diminish the efficacy of averaging; as the interrogation time increases, individual shifts become increasingly resolved, distorting the line shape of the multi-ion spectroscopy signal, and shifting the center away from the mean.   For $^{176}$Lu$^+$, the most significant problem arises from the coupling of the Coulomb interaction between ions to the EQ moment.  In a linear ion crystal, the shift is given by \cite{Itano2000}
\begin{equation}
h\Delta f_{Q,F} = \frac{C_F}{4}Q_j (3 \cos^2 \theta-1)\Theta(J) m \omega_z^2/e,
\label{QuadrupoleEq}
\end{equation}
where $e$ is the elementary charge, $m$ is the mass of the ion, $\theta$ is the angle between the applied dc magnetic field and the trap axis, $\{C_6,C_7,C_8\} = \{-2,5,-3\}/5$, and $\Theta(J) = 0.66\,\mathrm{a.u.}$ is the EQ moment for $^3D_1$ \cite{Safronova2018}. For a three-ion crystal, $\{Q_1,Q_2,Q_3\} = \{9,16,9\}/5$.  In general, the shift can be eliminated by aligning the field at an angle $\theta_0\approx 54.7^\circ$ to the trap axis.

Although it has been demonstrated that EMM can be mitigated in a well-designed linear Paul trap \cite{Keller2019}, the trap used in this work has axial end caps, which results in a significant rf-field along the trap axis.  Since this can partially mimic the form of Eq.~\ref{QuadrupoleEq}, EMM shifts must be characterised as well.  In general, EMM produces a second-order Doppler (SD) shift, $\Delta f_\mathrm{SD}$, which can be written
\begin{equation}
\frac{\Delta f_\mathrm{SD}}{f_0} = -\frac{1}{2}\sum_i\left( \frac{\omega_{i,rf} u_i}{c} \right)^2,
\label{SDEq1}
\end{equation}
where $f_0$ is the clock transition frequency, $u_i$ is the displacement of the ion along the $i^\mathrm{th}$ principle axis, and $\omega_{i,rf}$ is the confinement frequency from the rf trapping field along the same axis. An associated AC Stark shift, $\Delta f_{S,F}$ is also produced, which is proportional to the SD shift and can be written
\begin{equation}
\frac{\Delta f_{S,F}}{\Delta f_\mathrm{SD}} =\left(\frac{\Omega_\mathrm{rf}}{\Omega_\mathrm{s}}\right)^2 - \frac{C_F}{2}(3 \cos^2 \beta-1)\left(\frac{\Omega_\mathrm{rf}}{\Omega_\mathrm{T}}\right)^2,
\label{MMStarkShiftEq}
\end{equation}
where $\beta$ is the angle between the applied dc magnetic field and the local rf electric field.  The quantities $\Omega_\mathrm{s}$ and $\Omega_\mathrm{T}$ are determined, respectively, by the differential scalar and tensor polarizabilities of the clock transition.  The values reported in \cite{Arnold2019} give $\Omega_\mathrm{s}=2\pi\times 245^{+33}_{-24}\,\mathrm{MHz}$ and $\Omega_\mathrm{T}=2\pi\times 16.5(5)\,\mathrm{MHz}$.  Hence,  for the value of $\Omega_\mathrm{rf}$ used here, the first term of Eq.~\ref{MMStarkShiftEq} is $<0.006$ and can be neglected.  

Micromotion is compensated by adjusting dc voltages while monitoring EMM sidebands in three directions iteratively. In the $xz$ plane, the strength of the EMM is measured by a sideband-ratio method \cite{Berkeland1998} using stimulated-Raman transitions between $\ket{^3D_1,7,0}$ and $\ket{^3D_1,8,0}$ with two sets of 646-nm beams as shown in Fig. \ref{TrapAndBeamSetup}(b). The EMM modulation indices are determined to be 0.049(1) and 0.013(1) for $x$ and $z$ directions, respectively. To minimize EMM along the third direction, the scattering probability at the EMM sideband is measured using a near-resonant 646 nm beam with a $\bm{k}$-vector nominally aligned along $\frac{1}{\sqrt{2}}(\hat{y} - \hat{z})$ direction (Fig. \ref{TrapAndBeamSetup}(b)). This approach is limited by off-resonance scattering from $^3P_0$ with an estimated minimum deducible EMM modulation index of $\sim 0.1$. We put an upperbound on the fractional SD shift of $\sim -6.5\times 10^{-18}$ for a $^{176}$Lu$^+$ located at the EMM null point and this is limited by the sensitivity afforded by the 646-nm scattering method.

With EMM reasonably compensated for the center ion, only the $z$-component for the outer ions will significantly contribute to Eq.~\ref{SDEq1}.  With dc voltages set to zero, $\omega_{z,rf}$ is measured to be $\simeq 2\pi\times 28\,\mathrm{kHz}$, which gives a differential fractional frequency shift of $-2.2\times 10^{-17}$ between the middle and outer ions.  In addition, the micromotion will be predominately along the $z$-axis, such that $\beta\approx\theta$.  Consequently, suppressing the EQ shifts by setting $\theta$, will also suppress the EMM-induced ac stark shift.
\begin{figure*}
\includegraphics[width=0.95\linewidth]{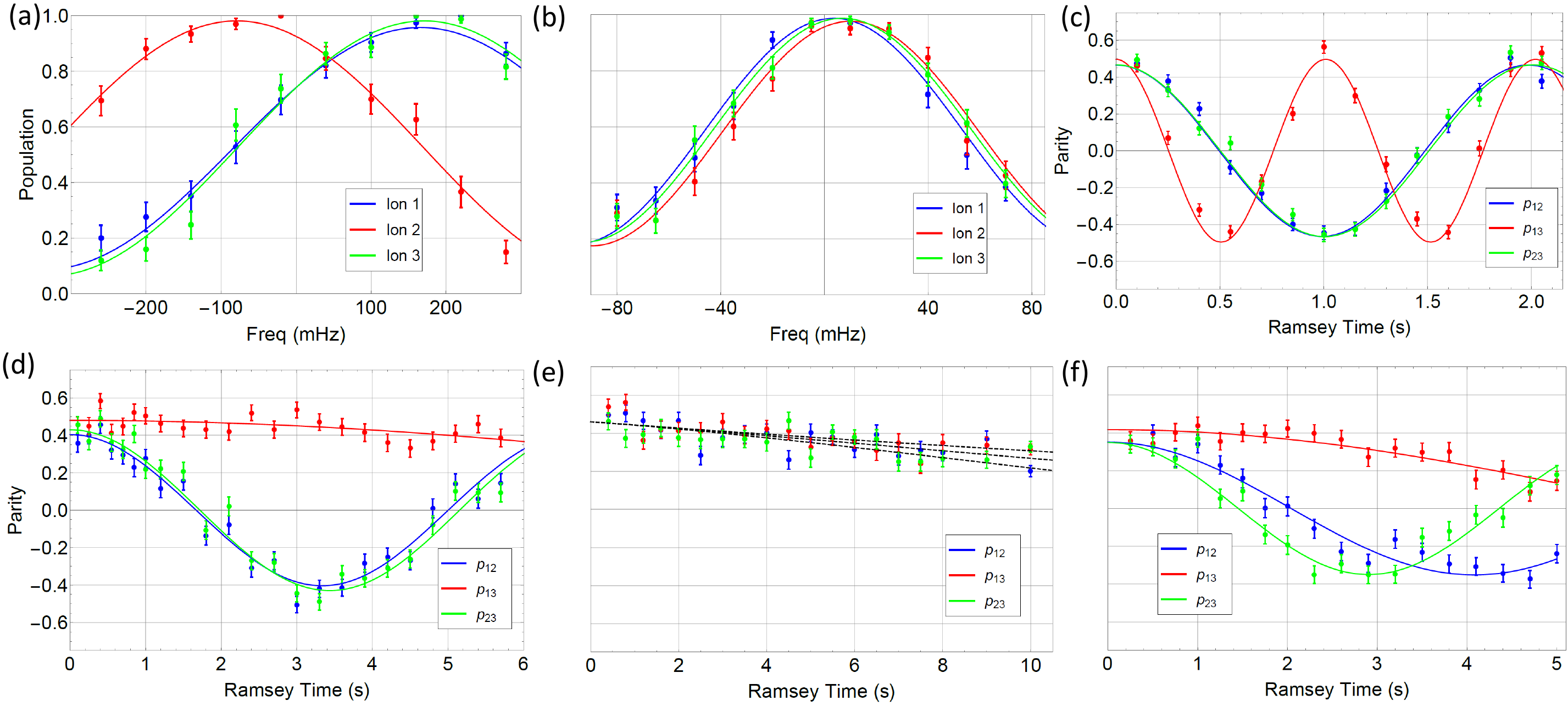}
\caption{(a) \& (b) Microwave Ramsey spectroscopy on the $\ket{{}^3D_1,7,0}\leftrightarrow\ket{{}^3D_1,8,0}$ transition with EMM minimized as described in the main text.  Frequencies on the horizontal axes are relative to an offset near to the hyperfine splitting of $\sim10.49\,\mathrm{GHz}$. For (a), a Ramsey time of $1\,\mathrm{s}$ is used with the B-field aligned approximately $20^{\circ}$ from $\theta_0$. Similarly (b) has a Ramsey time of $5\,\mathrm{s}$ with the B-field aligned to within $0.4^{\circ}$ of $\theta_0$;  (c) Correlation spectroscopy of the $\ket{{}^3D_1,7,0}\leftrightarrow\ket{{}^3D_1,8,1}$ transition. The oscillation period corresponds to a differential field of $\simeq 1.15\,\mathrm{nT}$ between the outer ions. (d)-(f). Correlation spectroscopy on the $\ket{{}^1S_0,7,1}\leftrightarrow\ket{{}^3D_1,7,0}$ transition:  (d) \& (e) have field alignments as for (a) \& (b); (f) has the B-field misaligned as in (a) and no EMM compensation.} 
\label{DataBinder}
\end{figure*}


To investigate inhomogeneities across the three-ion crystal, both optical and microwave Ramsey spectroscopy is used.  In either case, the probability for each ion to make a transition is  
\begin{equation}
P_i = \frac{1}{2}\left[ 1 + \cos \left(2\pi(f-f_{0,i})T\right ) \right],
\end{equation}
where $f$ is the frequency of the probe field, $f_{0,i}$ is the resonant transition of the $i^\mathrm{th}$ ion, and $T$ the Ramsey time. Correlation spectroscopy is also used to assess frequency comparisons between pairs of ions beyond the radiation field's coherence time \cite{Chwalla2007,Chou2011}. Following \cite{Chwalla2007,Chou2011}, the expectation value of the parity operator $p_{ij}=\langle\sigma_{z,i}\sigma_{z,j}\rangle$, when averaged over a uniform distribution of laser phases, is found to be
\begin{equation}
p_{ij} = \frac{p_c}{2} \cos\left[2\pi (f_{0,i} - f_{0,j})T\right],
\label{ParityInCoherentEq}
\end{equation}
where $p_c$ characterizes the loss of relative coherence between the two atomic oscillators under investigation. Note that correlation spectroscopy also rejects common-mode fluctuations influencing the atoms, such as that arising from ambient magnetic field noise.

The $\ket{{}^3D_1,7,0}\leftrightarrow\ket{{}^3D_1,8,0}$ transition has a very low sensitivity to magnetic fields, which facilitates long Ramsey times with high frequency resolution.  Moreover, when EMM is reasonably compensated, only EQ shifts given by Eq.~\ref{QuadrupoleEq} significantly contribute to a differential frequency shift between individual ions. The combination of high frequency resolution and the angular dependence of Eq.~\ref{QuadrupoleEq} facilitates accurate alignment of the B-field to suppress tensor shifts across the ion crystal.  With a misaligned magnetic field, individual ion signals for a $1\,\mathrm{s}$ long Ramsey interrogation are shown in Fig.~\ref{DataBinder}(a).  As expected, the outer two ions are shifted differentially with respect to the center ion.  The observed frequency difference of $252(10)\,\mathrm{mHz}$ and the theoretical quadrupole moment $\Theta=0.655\,\mathrm{a.u.}$ from Ref.~\cite{Safronova2018} implies an angular misalignment of $\sim20^\circ{}$.  

The frequency difference observed in Fig.~\ref{DataBinder}(a) is contributed from both states and is a factor of $8/5$ larger than for the optical transition to $\ket{{}^1S_0,7,1}$.  This is demonstrated by correlation spectroscopy of the $\ket{{}^1S_0,7,1}\leftrightarrow\ket{{}^3D_1,7,0}$ transition shown in Fig.~\ref{DataBinder}(d), which is taken under the same conditions as Fig.~\ref{DataBinder}(a).  As expected, the outer two ions maintain the same relative phase as function of interrogation time but accumulate a phase difference with respect to the middle ion.  The estimated frequency difference from this data is $148(3)\,\mathrm{mHz}$ consistent with the $252(10)\,\mathrm{mHz}$ observed in Fig.~\ref{DataBinder}(a).

Alignment of the field by reducing the observed frequency differences can be enhanced by increasing the Ramsey interrogation time. In Fig.~\ref{DataBinder}(b), the Ramsey time is increased to $5\,\mathrm{s}$ and the B-field adjusted to reduce the observed frequency differences to $4.9(1.6)\,\mathrm{mHz}$, which corresponds to $\sim10^{-17}$ of the  $^1S_0\leftrightarrow{}^3D_1$ optical transition. This would enable a clock interrogation duration of $\gtrsim10\,\mathrm{s}$ without significant broadening effects due to differential EQ shifts.  This is illustrated by the $10\,\mathrm{s}$ long optical correlation spectroscopy measurement shown in Fig.~\ref{DataBinder}(e).  Due to restricted optical access in our setup, we are unable to create a differential phase shift as demonstrated in Ref.~\cite{Chou2011}.  However, for the purposes of comparison, the optical correlation spectroscopy measurement shown in Fig.~\ref{DataBinder}(f) is taken with a misaligned magnetic field and without EMM compensation.  

Magnetic field gradients can also be assessed using microwave correlation spectroscopy on the $\ket{7,0}\leftrightarrow\ket{8,1}$ transition as shown in Fig. \ref{DataBinder}(c). As expected, a field gradient causes a phase accumulation between the outer ions at twice the rate relative to the inner ion.  Given the transition's field sensitivity of $875\,\mathrm{Hz/\mu T}$, the observed oscillation corresponds to a field gradient of $52\,\mathrm{\mu T/m}$ and a $\sim4\,\mathrm{mHz}$ frequency difference between the outer two ions for the $\ket{{}^1S_0,7,1}\leftrightarrow\ket{{}^3D_1,7,0}$ clock transition.

Dominant contributions to inhomogenous shifts of the $\ket{{}^1S_0,7,1}\leftrightarrow\ket{{}^3D_1,7,0}$ clock transition across the three ions consist of: a SD shift of $-7.8\,\mathrm{mHz}$ on the outer ions due to EMM, a $\pm 2\,\mathrm{mHz}$ shift on the outer ions due to magnetic field gradients,  a $3.1(1.1)\,\mathrm{mHz}$ differential shift between the middle and outer ions due to the EQ shift induced by neighbouring ions.  The sum total of these shifts cannot completely account for the overall decline of the parity signals seen in Fig.~\ref{DataBinder}(e).  We believe the drop in contrast to be caused by heating during the Ramsey zone, which similarly accounts for the loss of contrast in Fig.~\ref{DataBinder}(b).  Nevertheless, to allow for possible dephasing, we fit the data in Fig.~\ref{DataBinder}(e) to Eq.~\ref{ParityInCoherentEq} using $p_c=p_0^2\exp(-t/T)$, with $p_0$ and $T$ as free parameters and $f_{0,i}-f_{0,j}$ set to the measured values. A $\chi^2$-fit gives $p_0=0.96(2)$, consistent with the efficiency of state preparation, and $T=27(6)\,\mathrm{s}$, with a reduced chi-square, $\chi^2_\nu=0.84$. This level of dephasing would not significantly degrade the expected stability of $\sigma(\tau)=8.2\times10^{-17}/\sqrt{\tau}$ for a $10\,\mathrm{s}$ Ramsey interrogation with three unentangled ions.

The SD shift arising from EMM is a consequence of the end cap geometry of the trap. Better trap designs have shown shifts at the $10^{-19}$ level over millimeter length scales \cite{Keller2019}, and those results would be equally applicable to lutetium.  The shift due to magnetic field gradients could also be trivially compensated with an external coil.  This leaves only the EQ shifts, which give differential shifts at the $10^{-17}$ level.  For a $10\,\mathrm{s}$ Ramsey experiment, these would have minimal effect on the mean position of the center line and the efficacy of hyperfine averaging. From the EQ shifts measured here, we estimate that hyperfine averaging would suppress residual shifts to well below $10^{-20}$.  

An alternative strategy to mitigate shifts from EMM is to use the $^{1}S_0\leftrightarrow{}^3D_2$ transition.  This transition has a negative differential scalar polarizability, which allows EMM shifts to be suppressed by operating at a magic trap drive frequency \cite{Berkeland1998} estimated to be $\Omega_{RF}/2\pi \approx 32.9(1.3)$ MHz \cite{Arnold2018}.  Moreover, for this transition, EQ shifts are reduced due to the smaller $C_F$ coefficients for the $\ket{^3D_2,F,0}$ states.


In summary, it has been shown that inhomogeneous broadening in a crystal of three $^{176}$Lu$^+$ ions can be suppressed to the level of $10^{-17}$ of the clock transition.  This level of suppression is sufficient to allow hyperfine averaging to practically eliminate shifts due to inhomogeneous broadening for interrogation times up to a few tens of seconds and for a larger number of ions.  The approach demonstrated here will allow lutetium clock operation to extend to multiple ions and take full advantage of the superior clock properties offered by this atom \cite{Arnold2018}.  This work also conclusively demonstrates that a quadrupole moment does not preclude the possibility of a high accuracy multi-ion clock. 

\begin{acknowledgements}
This work is supported by the National Research Foundation, Prime Ministers Office, Singapore and the Ministry of Education, Singapore under the Research Centres of Excellence programme. This work is also supported by A*STAR SERC 2015 Public Sector Research Funding (PSF) Grant (SERC Project No: 1521200080). T. R. Tan acknowledges support from the Lee Kuan Yew post-doctoral fellowship.
\end{acknowledgements}
\bibliography{Multi-ion}

\begin{thebibliography}{10}

\bibitem{Ludlow2015}
Andrew~D Ludlow, Martin~M Boyd, Jun Ye, Ekkehard Peik, and Piet~O Schmidt.
\newblock Optical atomic clocks.
\newblock {\em Reviews of Modern Physics}, 87(2):637, 2015.

\bibitem{Nicholson2015}
TL~Nicholson, SL~Campbell, RB~Hutson, GE~Marti, BJ~Bloom, RL~McNally, Wei
  Zhang, MD~Barrett, MS~Safronova, GF~Strouse, et~al.
\newblock Systematic evaluation of an atomic clock at $2\times 10^{-18}$ total
  uncertainty.
\newblock {\em Nature communications}, 6:6896, 2015.

\bibitem{McGrew2018}
WF~McGrew, X~Zhang, RJ~Fasano, SA~Sch{\"a}ffer, K~Beloy, D~Nicolodi, RC~Brown,
  N~Hinkley, G~Milani, M~Schioppo, et~al.
\newblock Atomic clock performance beyond the geodetic limit.
\newblock {\em arXiv preprint arXiv:1807.11282}, 2018.

\bibitem{Chou2010}
Chin-wen Chou, DB~Hume, JCJ Koelemeij, David~J Wineland, and T~Rosenband.
\newblock Frequency comparison of two high-accuracy {Al}$^+$ optical clocks.
\newblock {\em Physical review letters}, 104(7):070802, 2010.

\bibitem{Huntemann2016}
N~Huntemann, C~Sanner, B~Lipphardt, Chr Tamm, and E~Peik.
\newblock Single-ion atomic clock with $3\times 10^{-18}$ systematic
  uncertainty.
\newblock {\em Physical review letters}, 116(6):063001, 2016.

\bibitem{Itano1993}
Wayne~M Itano, James~C Bergquist, John~J Bollinger, JM~Gilligan, DJ~Heinzen,
  FL~Moore, MG~Raizen, and David~J Wineland.
\newblock Quantum projection noise: Population fluctuations in two-level
  systems.
\newblock {\em Physical Review A}, 47(5):3554, 1993.

\bibitem{Wineland1994}
David~J Wineland, John~J Bollinger, Wayne~M Itano, and DJ~Heinzen.
\newblock Squeezed atomic states and projection noise in spectroscopy.
\newblock {\em Physical Review A}, 50(1):67, 1994.

\bibitem{Andre2004}
Axel Andr{\'e}, AS~S{\o}rensen, and MD~Lukin.
\newblock Stability of atomic clocks based on entangled atoms.
\newblock {\em Physical review letters}, 92(23):230801, 2004.

\bibitem{Bollinger1996}
John~J Bollinger, Wayne~M Itano, David~J Wineland, and DJ~Heinzen.
\newblock Optimal frequency measurements with maximally correlated states.
\newblock {\em Physical Review A}, 54(6):R4649, 1996.

\bibitem{Leibfried2004}
D~Leibfried, Murray~D Barrett, T~Schaetz, J~Britton, J~Chiaverini, Wayne~M
  Itano, John~D Jost, Christopher Langer, and David~J Wineland.
\newblock Toward heisenberg-limited spectroscopy with multiparticle entangled
  states.
\newblock {\em Science}, 304(5676):1476--1478, 2004.

\bibitem{Shaniv2018}
Ravid Shaniv, Tom Manovitz, Yotam Shapira, Nitzan Akerman, and Roee Ozeri.
\newblock Toward heisenberg-limited rabi spectroscopy.
\newblock {\em Physical review letters}, 120(24):243603, 2018.

\bibitem{Borregaard2013}
Johannes Borregaard and Anders~S{\o}ndberg S{\o}rensen.
\newblock Efficient atomic clocks operated with several atomic ensembles.
\newblock {\em Physical review letters}, 111(9):090802, 2013.

\bibitem{Rosenband2013arXiv}
T~Rosenband and DR~Leibrandt.
\newblock Exponential scaling of clock stability with atom number.
\newblock {\em arXiv preprint arXiv:1303.6357}, 2013.

\bibitem{Keller2019}
J~Keller, D~Kalincev, T~Burgermeister, AP~Kulosa, A~Didier, T~Nordmann,
  J~Kiethe, and TE~Mehlst{\"a}ubler.
\newblock Probing time dilation in coulomb crystals in a high-precision ion
  trap.
\newblock {\em Physical Review Applied}, 11(1):011002, 2019.

\bibitem{Berkeland1998}
DJ~Berkeland, JD~Miller, James~C Bergquist, Wayne~M Itano, and David~J
  Wineland.
\newblock Minimization of ion micromotion in a paul trap.
\newblock {\em Journal of applied physics}, 83(10):5025--5033, 1998.

\bibitem{Arnold2015}
Kyle Arnold, Elnur Hajiyev, Eduardo Paez, Chern~Hui Lee, MD~Barrett, and John
  Bollinger.
\newblock Prospects for atomic clocks based on large ion crystals.
\newblock {\em Physical Review A}, 92(3):032108, 2015.

\bibitem{Shaniv2018arXiv}
Ravid Shaniv, Nitzan Akerman, Tom Manovitz, Yotam Shapira, and Roee Ozeri.
\newblock Quadrupole shift cancellation using dynamic decoupling.
\newblock {\em arXiv preprint arXiv:1808.10727}, 2018.

\bibitem{aharon2018robust}
Nati Aharon, Nicolas Spethmann, Ian~D Leroux, Piet~O Schmidt, and Alex Retzker.
\newblock Robust optical clock transitions in trapped ions.
\newblock {\em arXiv preprint arXiv:1811.06732}, 2018.

\bibitem{Barrett2015}
M.~D. Barrett.
\newblock Developing a field independent frequency reference.
\newblock {\em New Journal of Physics}, 17(5):053024, 2015.

\bibitem{Kaewuam2017}
Rattakorn Kaewuam, Arpan Roy, Ting~Rei Tan, KJ~Arnold, and MD~Barrett.
\newblock Laser spectroscopy of $^{176}${Lu}$^+$.
\newblock {\em Journal of Modern Optics}, 65(5-6):592--601, 2018.

\bibitem{Kaewuam2019}
R~Kaewuam, TR~Tan, KJ~Arnold, and MD~Barrett.
\newblock Spectroscopy of the $^1{S}_0$ to $^1{D}_2$ clock transition in
  $^{176}${Lu}$^+$.
\newblock {\em Phys. Rev. A}, 99:022514, 2019.

\bibitem{Itano2000}
Wayne~M Itano.
\newblock External-field shifts of the $^{199}${Hg}$^+$ optical frequency
  standard.
\newblock {\em Jour. Res. of the Nat. Inst. of Stan. and Tech.}, 105(6):829,
  2000.

\bibitem{Safronova2018}
SG~Porsev, UI~Safronova, and MS~Safronova.
\newblock Clock-related properties of {Lu}$^+$.
\newblock {\em Physical Review A}, 98(2):022509, 2018.
\newblock The quadrupole moments given use a nuclear physics convention, which
  includes an extra factor of two. For consistency with notation used here and
  in \cite{Itano2000}, they must be multiplied by $-1/2$.

\bibitem{Arnold2019}
KJ~Arnold, R~Kaewuam, TR~Tan, SG~Porsev, MS~Safronova, and MD~Barrett.
\newblock Dynamic polarizability measurements with $^{176}${Lu}$^+$.
\newblock {\em Physical Review A}, 99(1):012510, 2019.

\bibitem{Chwalla2007}
M~Chwalla, K~Kim, T~Monz, P~Schindler, M~Riebe, CF~Roos, and R~Blatt.
\newblock Precision spectroscopy with two correlated atoms.
\newblock {\em Applied Physics B}, 89(4):483--488, 2007.

\bibitem{Chou2011}
Chin-Wen Chou, DB~Hume, Michael~J Thorpe, David~J Wineland, and T~Rosenband.
\newblock Quantum coherence between two atoms beyond ${Q}=10^{15}$.
\newblock {\em Physical review letters}, 106(16):160801, 2011.

\bibitem{Arnold2018}
Kyle~J Arnold, Rattakorn Kaewuam, Arpan Roy, Ting~Rei Tan, and Murray~D
  Barrett.
\newblock Blackbody radiation shift assessment for a lutetium ion clock.
\newblock {\em Nature communications}, 9:1650, 2018.

\end{thebibliography}
\bibliographystyle{unsrt}
\end{document}